\documentclass[11pt,a4paper]{article}
\usepackage[margin=2.5cm]{geometry}
\usepackage{amsmath,amssymb,amsfonts}
\usepackage{times}
\usepackage{setspace}
\usepackage[numbers,sort&compress]{natbib}
\usepackage[hidelinks]{hyperref}
\usepackage{graphicx}
\usepackage{subcaption}
\usepackage{microtype}
\usepackage{titlesec}
\usepackage{indentfirst}

\titleformat{\section}{\large\bfseries}{\thesection.}{0.5em}{}
\titleformat{\subsection}{\normalsize\bfseries}{\thesubsection}{0.5em}{}

\title{\textbf{From a Long-standing Prediction to a New Family of Hadrons}}

\author{%
  Xining Wang\textsuperscript{1},
  Kai Yi\textsuperscript{1,2,*}\\[0.8em]
  \normalsize
  Ministry of Education Key Laboratory of NSLSCS,\\
  Institute of Physics Frontiers and Interdisciplinary Sciences\\
  \textsuperscript{1} School of Physics and Technology, Nanjing Normal University,
  Nanjing 210023, China\\
  \textsuperscript{2} Department of Physics, Tsinghua University,
  Beijing 100084, China\\[0.6em]
  \textit{Perspective}\\[0.4em]
  \textit{For the IOP Focus Collection ``A Century of Physics at Tsinghua:
  Legacy and Frontiers''}\\[0.6em]
  $^{*}$Author to whom any correspondence should be addressed.\\
  E-mail: \texttt{kai.yi@njnu.edu.cn}
}

\date{}

\begin{document}
\maketitle
\thispagestyle{empty}

\begin{abstract}
\noindent
Fully charmed tetraquarks have evolved from a long-standing theoretical prediction into one of the most active frontiers in hadron spectroscopy. This article reviews their development from early theoretical studies and experimental searches to the recent breakthroughs at the Large Hadron Collider, with particular emphasis on the contributions of the Nanjing Normal University (NNU) and Tsinghua University (THU) teams at the CMS experiment. We summarize the discovery of the fully charmed tetraquark family, the establishment of interference among resonances, and the first determination of their quantum numbers. These advances have transformed the search for a single exotic resonance into the study of a new family of hadrons and opened a promising avenue for exploring the nonperturbative dynamics of Quantum Chromodynamics.
\end{abstract}

\noindent\textbf{Keywords:} fully charmed tetraquarks, CMS, LHC, quantum numbers, hadron spectroscopy

\section{Introduction}

One of the central questions in hadron physics is deceptively simple: what combinations of quarks can exist in nature? While the quark model successfully describes hadrons as quark--antiquark mesons and three-quark baryons, Quantum Chromodynamics (QCD) allows a much richer spectrum of color-singlet states. The discoveries of the $X(3872)$~\cite{Belle:2003nnu} and many subsequent XYZ particles established multiquark hadrons as an important frontier of modern hadron spectroscopy~\cite{Chen:2016qju,Ali:2017jda}, but their internal structures remain a subject of active debate.

Among these exotic candidates, fully heavy tetraquarks are particularly attractive. The all-charm system, $cc\bar{c}\bar{c}$, contains no light valence quarks and is therefore theoretically cleaner than conventional XYZ states. Experimentally, its expected decay into $J/\psi\,J/\psi$ provides a remarkably clean signature at the Large Hadron Collider (LHC), where abundant prompt $J/\psi$ production and excellent dimuon reconstruction offer unique discovery potential.

The possibility of fully heavy tetraquarks was proposed more than four decades ago~\cite{Iwasaki:1975pv,Chao:1980dv,Ader:1981eb,Zouzou:1986qh}, yet convincing experimental evidence remained absent until the LHC era. Only with the large datasets accumulated by the LHC has it become possible to transform long-standing theoretical predictions into quantitative experimental measurements.

This article reviews that development, with particular emphasis on the contributions of the Nanjing Normal University (NNU) and Tsinghua University (THU) teams at the CMS experiment~\cite{Zhu:2024swp}. From exploratory CMS Run~1 studies to the discovery of the all-charm tetraquark family and the first determination of its quantum numbers, these efforts have helped transform the field from the search for a single resonance into an emerging spectroscopy of fully heavy hadrons.

\section{Early History: Before the Discovery}

\subsection{Experimental Clues}

The idea of fully heavy tetraquarks was motivated not only by QCD but also by several intriguing experimental observations. Following the discovery of the $J/\psi$, a Fermilab experiment led by Lederman reported a dilepton enhancement near $6\,\mathrm{GeV}$~\cite{Hom:1976wq}. Iwasaki then suggested that this putative resonance was the $cc\bar{c}\bar{c}$ state he had predicted the previous year~\cite{Iwasaki:1975pv,Iwasaki:1976uh}. Further data soon showed that the bump was a statistical fluctuation, but the theoretical idea survived.

\begin{figure}[htbp]
\centering
\includegraphics[width=0.49\textwidth]{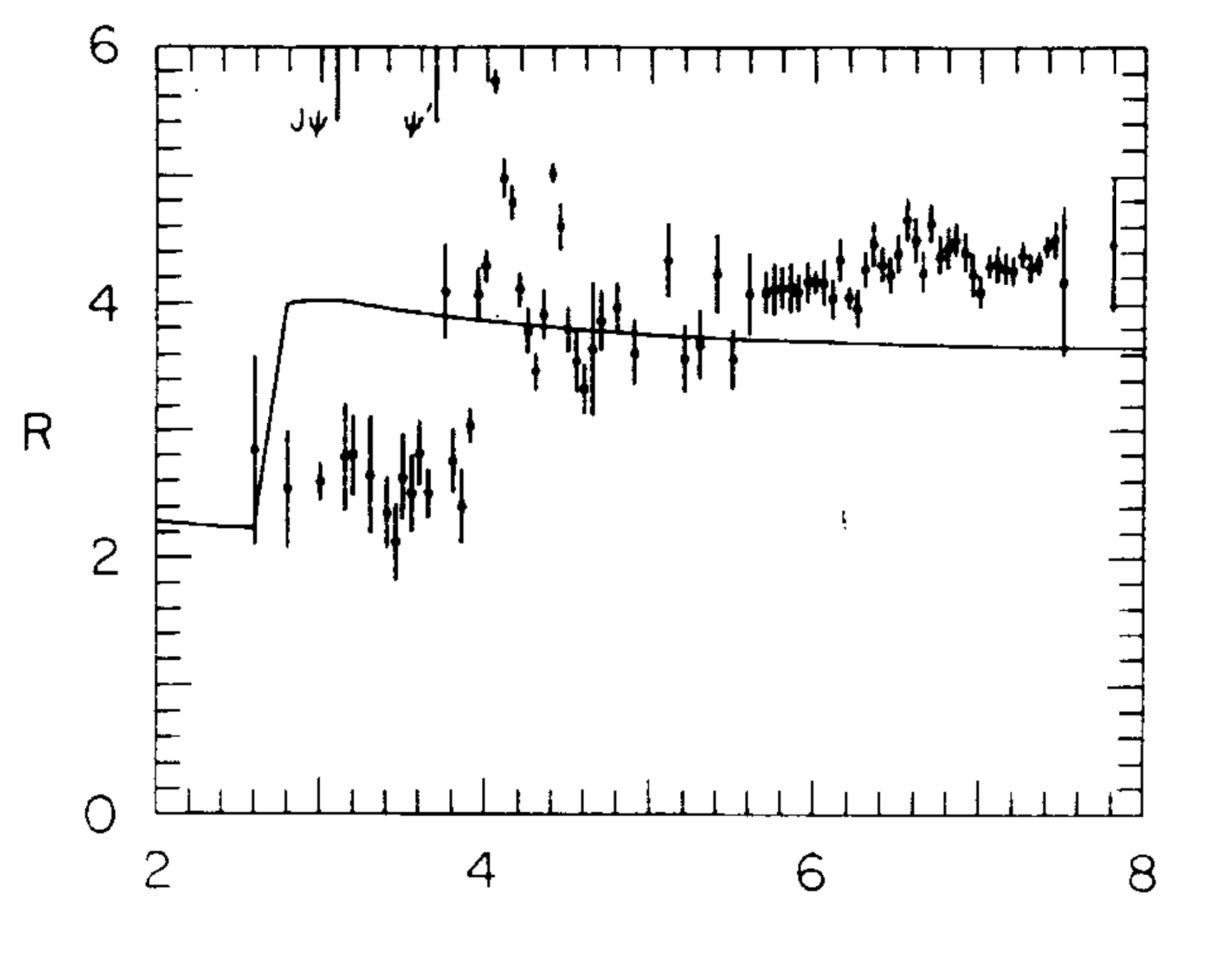}
\caption{The ratio $R$ as a function of the center-of-mass energy $\sqrt{s}$, comparing SLAC--LBL data~\cite{Augustin:1975yq,Siegrist:1976br} with a QCD prediction (solid curve)~\cite{Barnett:1980sm}. Possible resonance-like structures in the $6$--$7\,\mathrm{GeV}$ region motivated early interpretations in terms of fully charmed tetraquark production~\cite{Chao:1980dv}.}
\label{fig:R-ratio}
\end{figure}

A related clue came from $e^{+}e^{-}$ annihilation. Experiments measured the hadronic cross-section ratio
\begin{equation*}
R = \sigma(e^{+}e^{-} \rightarrow \text{hadrons})/\sigma(e^{+}e^{-} \rightarrow \mu^{+}\mu^{-}),
\end{equation*}
which compares the rate of hadron production in $e^{+}e^{-}$ annihilation with that of muon pairs. This ratio is sensitive to new quark flavors, which raise the hadronic yield, and to resonances, which appear as peaks. Measurements at SLAC and LBL revealed unexplained structures in the high-mass charmonium region~\cite{Augustin:1975yq,Siegrist:1976br}. Figure~\ref{fig:R-ratio} shows an early comparison of the measured $R$ with the QCD expectation, in which possible resonance-like structures appear above the charm threshold and in the $6$--$7\,\mathrm{GeV}$ region~\cite{Barnett:1980sm,Chao:1980dv}. Although these anomalies were never interpreted conclusively, they further stimulated interest in unconventional multiquark configurations~\cite{Chao:1980dv}.

\subsection{Theoretical Predictions}

Beginning in the mid-1970s and early 1980s, Iwasaki~\cite{Iwasaki:1975pv,Iwasaki:1976uh}, Chao~\cite{Chao:1980dv}, and Ader, Richard, and Taxil~\cite{Ader:1981eb} explored the possibility of bound states containing only heavy quarks. Their pioneering work, including early diquark--antidiquark studies, established the early theoretical basis for fully heavy tetraquarks.

Over the following decades, constituent quark models, diquark--antidiquark models, effective field theories, and lattice QCD all investigated the $cc\bar{c}\bar{c}$ system. Chinese and Japanese groups played particularly important roles in developing these studies, predicting a rich spectrum of states between approximately $6.2$ and $7.2\,\mathrm{GeV}$~\cite{Wu:2016vtq,Liu:2019zuc,Debastiani:2017msn,Bedolla:2019uqp}. Although quantitative predictions varied, most models agreed that compact fully charmed tetraquarks could exist and would decay predominantly into $J/\psi\,J/\psi$, providing a clean experimental signature.

\subsection{Four Decades of Searching}

\begin{figure}[htbp]
\centering
\subcaptionbox{Dimuon mass combinations\label{fig:na3-2d}}{%
  \includegraphics[height=5.44cm]{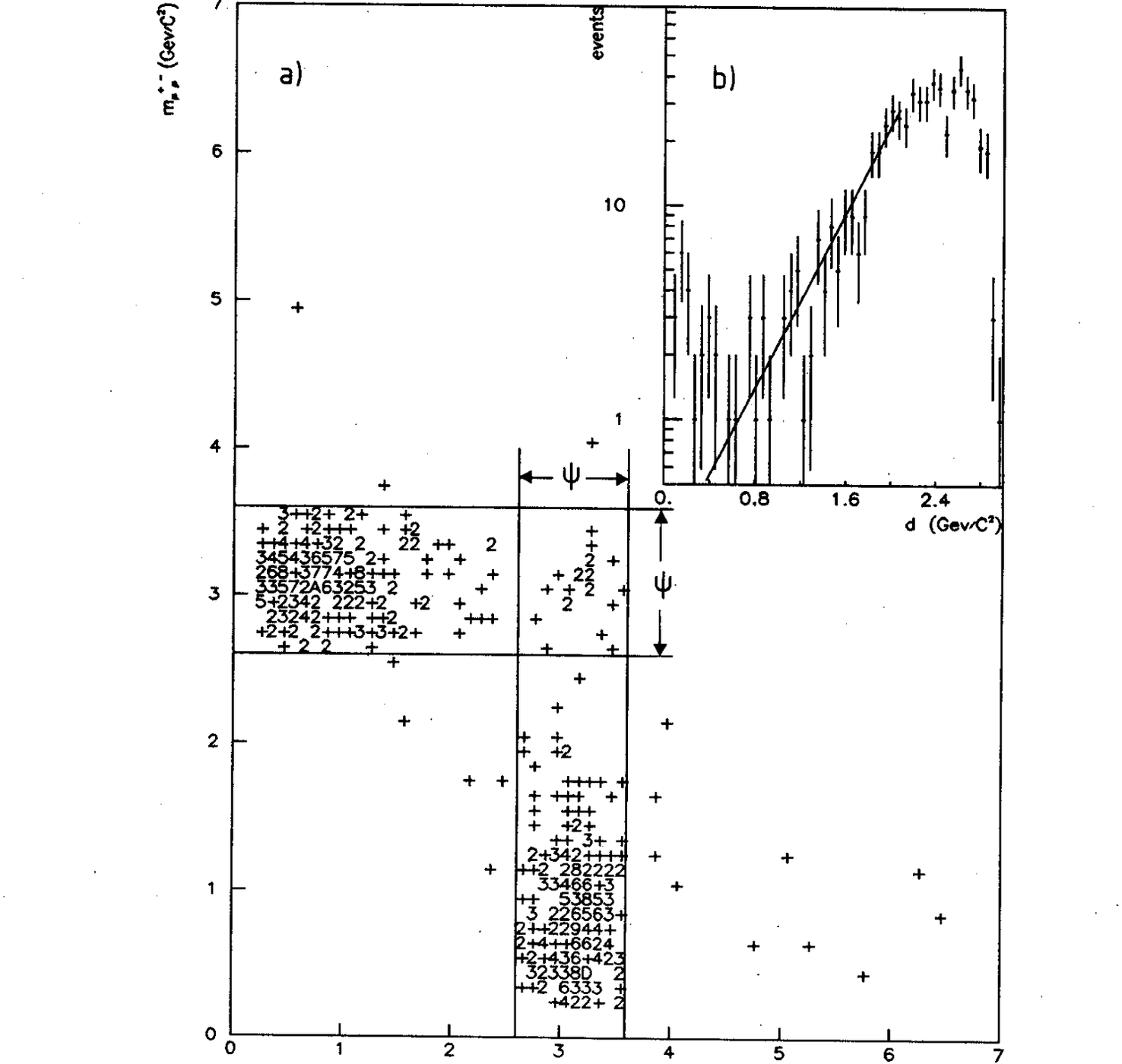}}%
\hspace{0.5em}%
\subcaptionbox{$J/\psi\,J/\psi$ mass spectrum\label{fig:na3-mass}}{%
  \includegraphics[height=5.44cm]{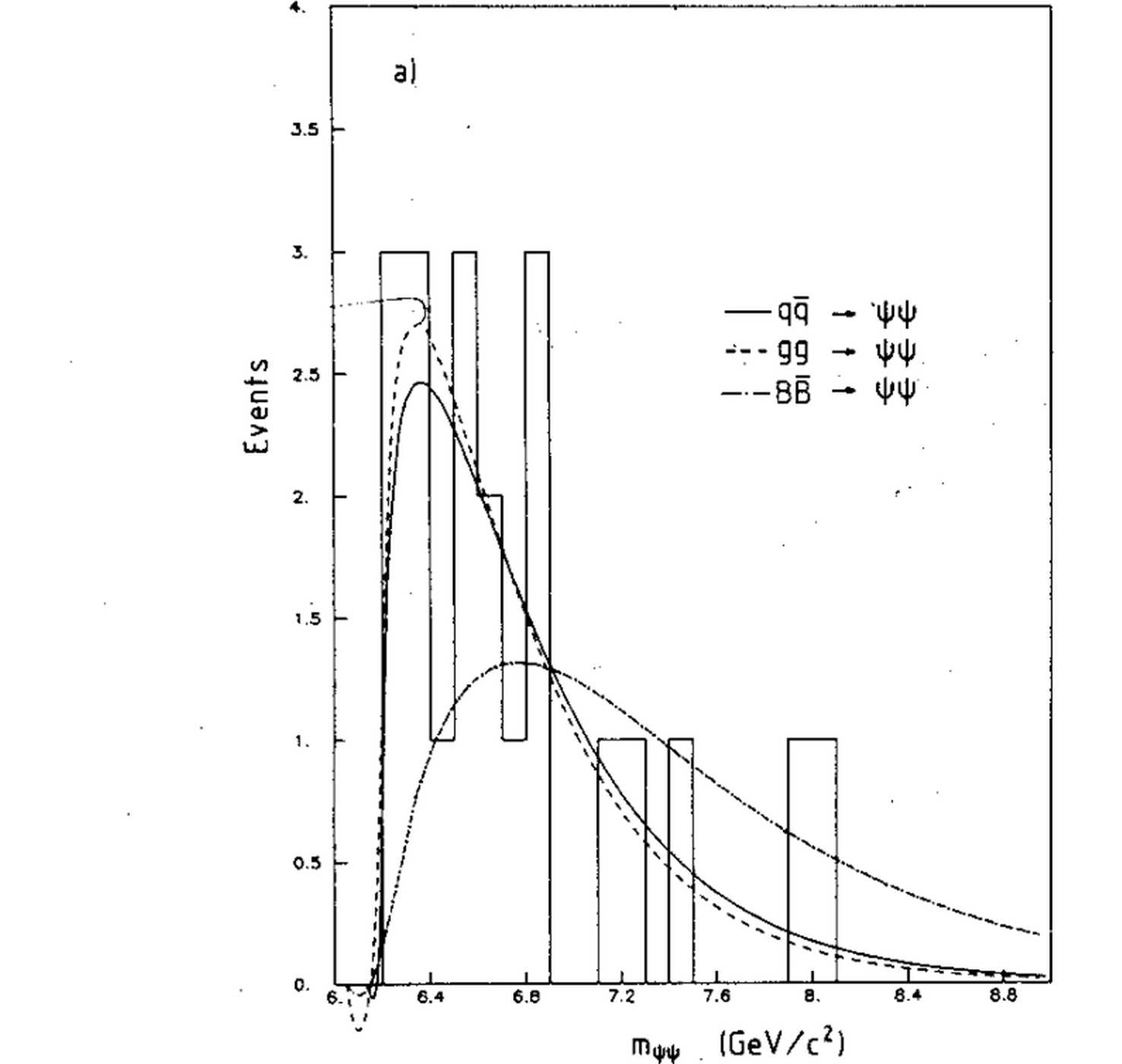}}
\caption{Early $J/\psi\,J/\psi$ results from the NA3 experiment at the CERN SPS with 400~GeV protons: (a)~distribution of the two dimuon mass combinations in four-muon events, with the $J/\psi$ bands indicated; (b)~$J/\psi\,J/\psi$ invariant-mass spectrum compared with QCD expectations for $q\bar{q}$, $gg$, and $B\bar{B}$ production mechanisms~\cite{NA3:1985rmd}.}
\label{fig:na3}
\end{figure}

Despite growing theoretical support, experimental confirmation remained elusive for nearly four decades. Searches at $e^{+}e^{-}$ colliders were limited by low production rates and insufficient statistics~\cite{Boyarski:1975yt,Aschman:1977kz}. The first intriguing indication in the expected discovery channel came from the NA3 experiment at the CERN SPS, which reported evidence for $J/\psi\,J/\psi$ production in fixed-target $\pi^{-}$ and proton interactions~\cite{NA3:1982qlq,NA3:1985rmd}. Figure~\ref{fig:na3} shows the two-dimensional distribution of the dimuon mass combinations and the $J/\psi\,J/\psi$ invariant-mass spectrum from the 400~GeV proton data. However, the available statistics and mass resolution were insufficient to establish a resonant structure.

By the mid-2010s, fully heavy tetraquarks remained an attractive theoretical prediction without convincing experimental confirmation. The breakthrough came at the LHC: with its enormous prompt $J/\psi$ production rate, excellent dimuon reconstruction, and high trigger efficiency, it became the ideal facility to search for resonances decaying to $J/\psi\,J/\psi$, setting the stage for systematic studies by CMS, ATLAS, and LHCb.

\section{Early CMS Searches}

The LHC was not designed as a hadron spectroscopy facility. Its primary goal was to explore the TeV energy scale, highlighted by the discovery of the Higgs boson in 2012~\cite{ATLAS:2012yve,CMS:2012qbp}. However, its enormous heavy-quark production, including billions of prompt $J/\psi$ mesons annually, has opened unprecedented opportunities for studying rare charmonium and multiquark states~\cite{Zhao:2025sym}.

CMS is well suited to this dimuon spectroscopy~\cite{CMS:2008zzk}. A silicon tracker in a 3.8~T solenoid, a muon system covering a wide central rapidity range, and an efficient dimuon trigger provide good $J/\psi \rightarrow \mu^{+}\mu^{-}$ mass resolution over a broad kinematic range. These features enable searches for rare resonances decaying into $J/\psi$ pairs, complementary to LHCb's forward acceptance at lower transverse momentum and to the similar central coverage of ATLAS.

\subsection{The First Exploratory Studies}

The possibility of searching for fully charmed tetraquarks with CMS data emerged soon after Run~1. Motivated by CDF and CMS exotic hadron studies in the $J/\psi\phi$ system~\cite{CDF:2009esx,CMS:2013qig,Zhou:2025nyj}, exploratory analyses using CMS 7 and 8~TeV data examined four-muon final states, including $J/\psi\,J/\psi$, $J/\psi\mu\mu$, and $\Upsilon\mu\mu$ channels~\cite{Yi:2018fxo,Durgut:2018thesis,Durgut:2018aps}. As emphasized in an earlier review~\cite{Yi:2013iok}, ``It will also be interesting to search for vector-vector structures composed entirely of c and b quarks near threshold because they may offer simpler systems to model theoretically.''

\begin{figure}[htbp]
\centering
\subcaptionbox{$\Upsilon(1S)\mu^{+}\mu^{-}$\label{fig:cms-upsilon-mumu}}{%
  \includegraphics[height=6.24cm]{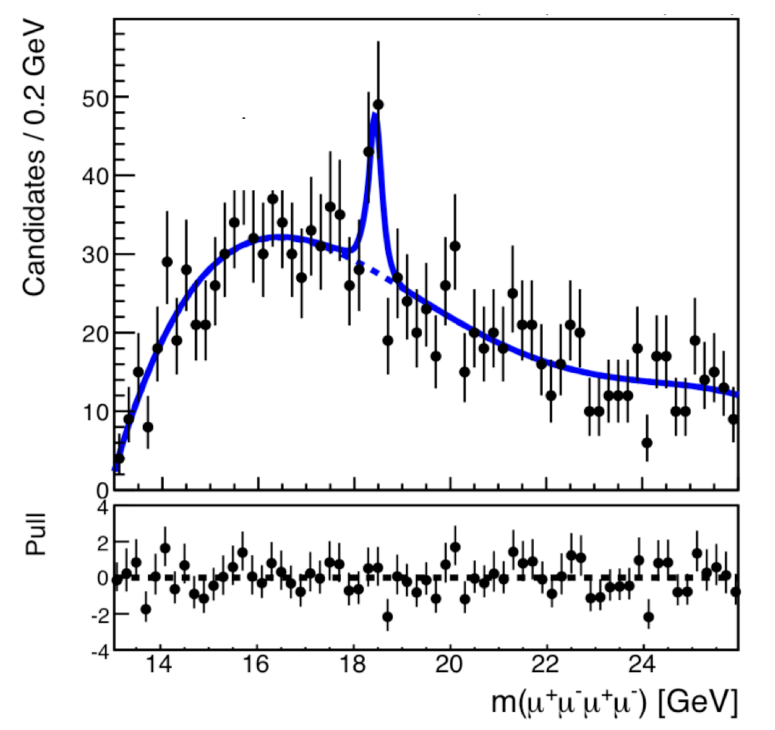}}%
\hspace{0.6em}%
\subcaptionbox{$\Upsilon(1S)e^{+}e^{-}$\label{fig:cms-upsilon-ee}}{%
  \includegraphics[height=6.24cm]{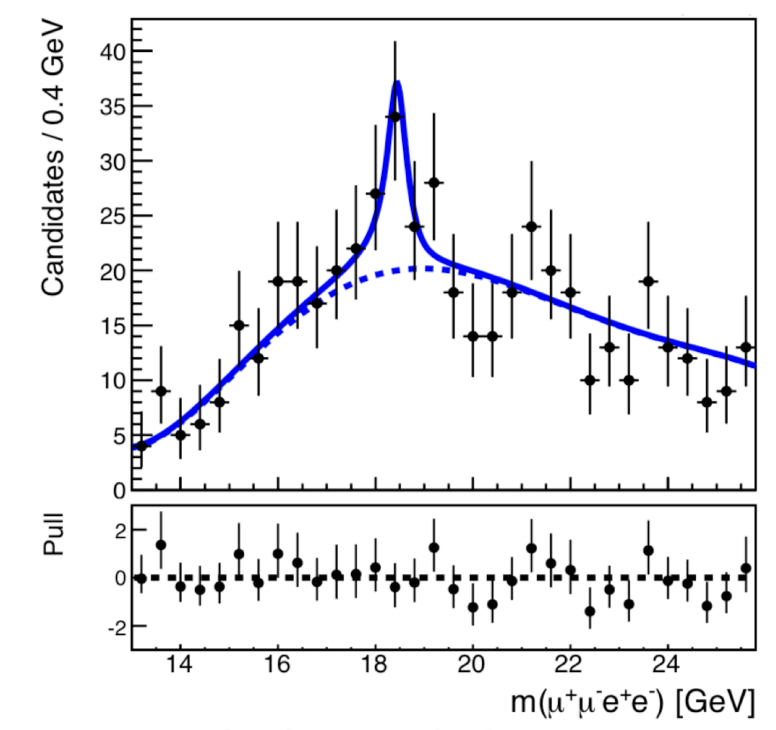}}
\caption{Invariant-mass spectra of $\Upsilon(1S)\ell^{+}\ell^{-}$ candidates using CMS data at 7 and 8~TeV, for (a)~$\Upsilon(1S)\mu^{+}\mu^{-}$ and (b)~$\Upsilon(1S)e^{+}e^{-}$~\cite{Durgut:2018thesis,Durgut:2018aps}.}
\label{fig:cms-upsilon}
\end{figure}

Figure~\ref{fig:cms-upsilon} shows the $\Upsilon(1S)\mu^{+}\mu^{-}$ and $\Upsilon(1S)e^{+}e^{-}$ invariant-mass spectra from these studies using CMS data. Both channels display a narrow enhancement near $18.5\,\mathrm{GeV}$, below the $\Upsilon(1S)\Upsilon(1S)$ threshold~\cite{Durgut:2018thesis,Durgut:2018aps}. This was not a discovery. The analysis was never a CMS Collaboration publication: it appeared only in a thesis and a conference talk, the combined local (global) significance was $4.9\sigma$ ($3.6\sigma$), and the result was sensitive to selection and background modeling. ATLAS later reported a similar excess near $18\,\mathrm{GeV}$ in 8~TeV $\Upsilon(1S)\mu^{+}\mu^{-}$ data, but it was not confirmed in their 13~TeV sample~\cite{ATLAS:2023nmu}. Published searches by LHCb~\cite{LHCb:2018hoe} and CMS~\cite{CMS:2020qvv} found no significant resonance in this region and set upper limits. The episode still mattered: it developed the four-lepton reconstruction and fitting methods later used for $J/\psi\,J/\psi$ spectroscopy.

\begin{figure}[htbp]
\centering
\subcaptionbox{2012 data\label{fig:cms-run1-2012}}{%
  \includegraphics[height=4.8cm]{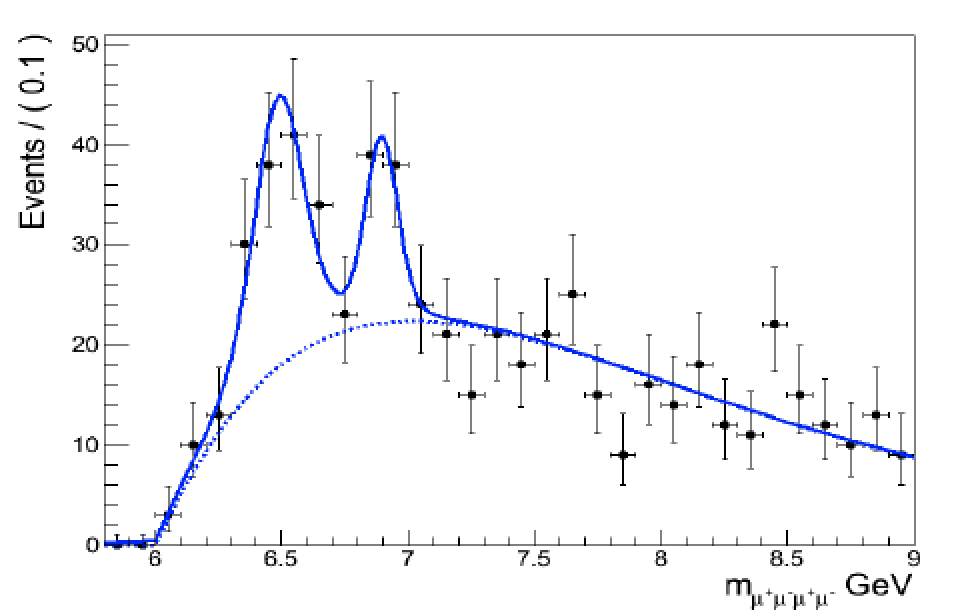}}%
\hspace{0.8em}%
\subcaptionbox{2011 data\label{fig:cms-run1-2011}}{%
  \includegraphics[height=4.8cm]{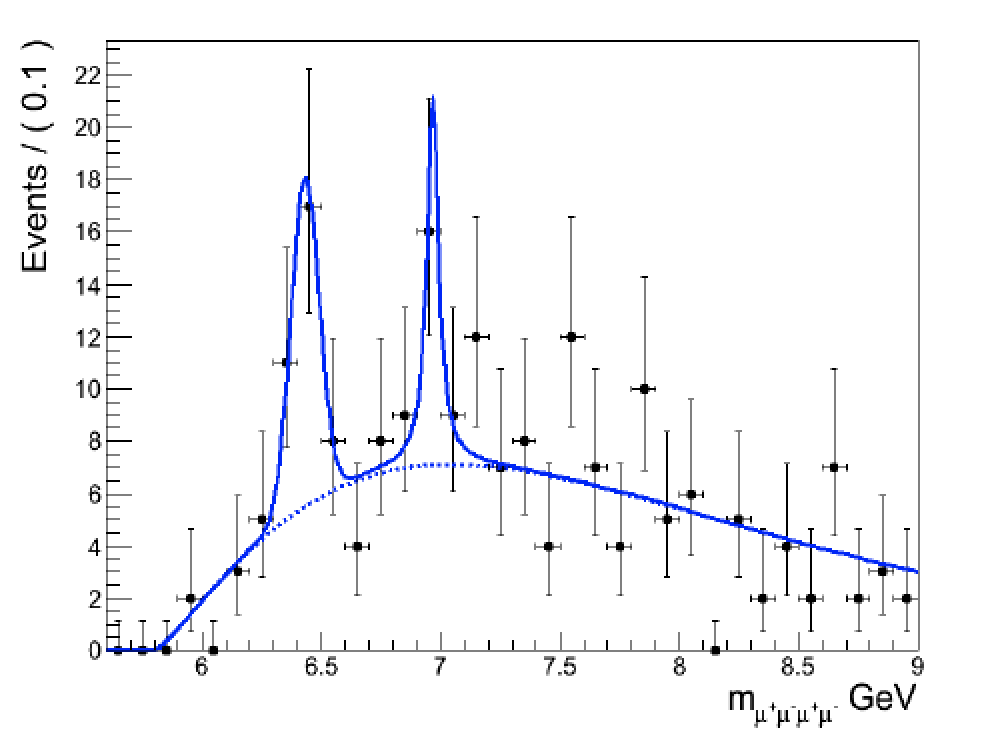}}
\caption{Four-muon invariant-mass spectra using CMS data collected in (a)~2012 and (b)~2011~\cite{Wang:2026phd}.}
\label{fig:cms-run1}
\end{figure}

The same studies also examined the $J/\psi\,J/\psi$ channel using CMS Run~1 data, as documented in CMS internal notes and later summarized in a Ph.D.\ thesis~\cite{Wang:2026phd} and a review article~\cite{Zhu:2024swp}. Figure~\ref{fig:cms-run1} shows the four-muon invariant-mass spectra using CMS data collected in 2012 and 2011, in which two enhancements near $6.5$ and $7.0\,\mathrm{GeV}$ appeared as hints.

The Run~1 samples were too small to establish whether these enhancements were genuine resonances. CMS resumed the search in 2019 with the much larger Run~2 datasets.

\subsection{Restarting the Program}

The effort was revived in October 2019 and evolved from the exploratory Run~1 studies into a long-term spectroscopy program. The goal extended beyond a simple resonance search to establishing a comprehensive framework for determining whether fully charmed tetraquarks exist, how many states are present, and what their properties are.

To that end, the analysis had to describe a $J/\psi\,J/\psi$ spectrum shaped simultaneously by prompt production, nonprompt backgrounds, threshold structures, detector resolution, and possible interference among overlapping states. Developing fitting strategies, statistical methods, and physics models that could accommodate these ingredients together was a central focus of the 2019--2020 effort.
Based on the Run~1 studies (Fig.~\ref{fig:cms-run1}), three blinded search regions were initially planned: $6.3$--$6.6$, $6.8$--$7.1$, and $7.2$--$7.8\,\mathrm{GeV}$. The first two covered the visible enhancements near $6.5$ and $7.0\,\mathrm{GeV}$, while the third targeted possible higher-mass states.

\section{From Exploration to Discovery}

\subsection{The LHCb Discovery}

\begin{figure}[htbp]
\centering
\begin{subfigure}[t]{0.48\textwidth}
\centering
\includegraphics[width=\linewidth]{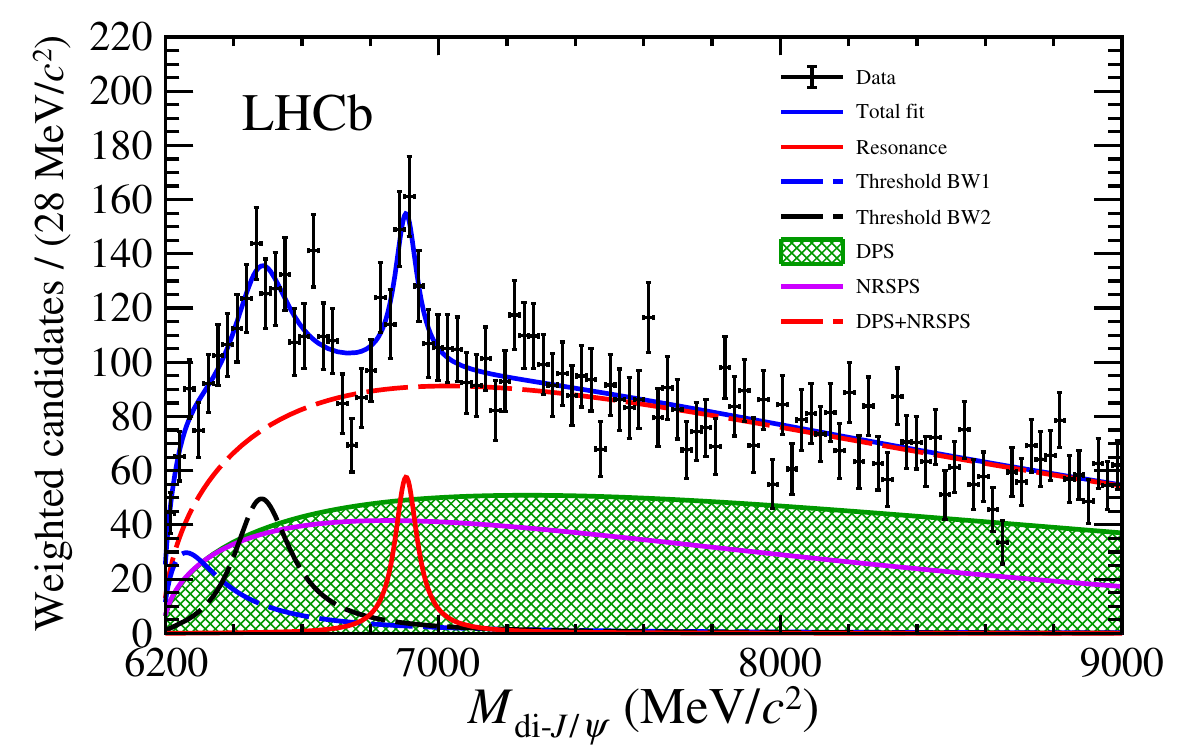}
\caption{Model~I (no interference)}
\label{fig:lhcb-nointerf}
\end{subfigure}\hfill
\begin{subfigure}[t]{0.48\textwidth}
\centering
\includegraphics[width=\linewidth]{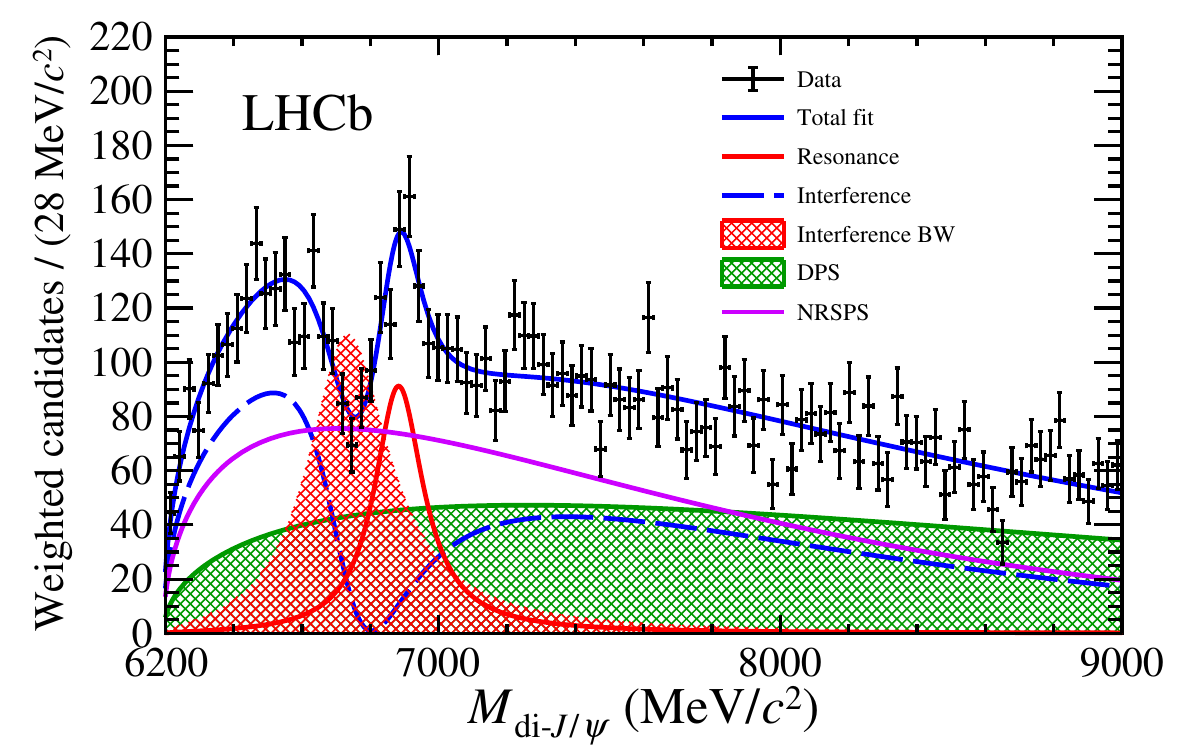}
\caption{Model~II (auxiliary BW interfering with the continuum)}
\label{fig:lhcb-interf}
\end{subfigure}
\caption{$J/\psi\,J/\psi$ invariant-mass spectra measured by LHCb, showing a significant enhancement near $6.9\,\mathrm{GeV}$. Fits are shown for (a)~Model~I, with no interference, and (b)~Model~II, in which an auxiliary Breit--Wigner amplitude interferes with the nonresonant continuum. This $X(6900)$ is added incoherently in both models~\cite{LHCb:2020bwg}.}
\label{fig:lhcb}
\end{figure}

While the CMS analysis was in progress, the LHCb Collaboration reported the first observation of a fully charmed tetraquark candidate in 2020~\cite{LHCb:2020bwg}. Using Run~1 and Run~2 $pp$ collision data, LHCb observed an enhancement around $6.9\,\mathrm{GeV}$ (Fig.~\ref{fig:lhcb}) with a local significance exceeding $5\sigma$. LHCb modeled the spectrum in two scenarios: a fit without interference, and a fit in which an auxiliary Breit--Wigner amplitude interferes with the nonresonant continuum while the $X(6900)$ is added incoherently~\cite{LHCb:2020bwg}.

After LHCb reported the enhancement near $6.9\,\mathrm{GeV}$, CMS merged the three blinded regions into a single broad blinded window, allowing an unbiased investigation of the complete $J/\psi\,J/\psi$ mass spectrum.

\subsection{Complementary Strategies at CMS}

CMS and LHCb differed not only in detector acceptance and resolution, but also in their analysis strategies for modeling the $J/\psi\,J/\psi$ spectrum. CMS benefits from large acceptance at high transverse momentum and excellent mass resolution in the central region, providing complementary kinematic coverage. CMS adopted a different modeling strategy: a global likelihood fit to the full spectrum that compares hypotheses with different numbers of resonances, both without interference and with interference among the overlapping resonant amplitudes themselves. Interference among nearby states can substantially alter the observed resonance line shapes.

\subsection{Discovery of the All-Charm Tetraquark Family}

\begin{figure}[htbp]
\centering
\begin{subfigure}[t]{0.48\textwidth}
\centering
\includegraphics[width=\linewidth]{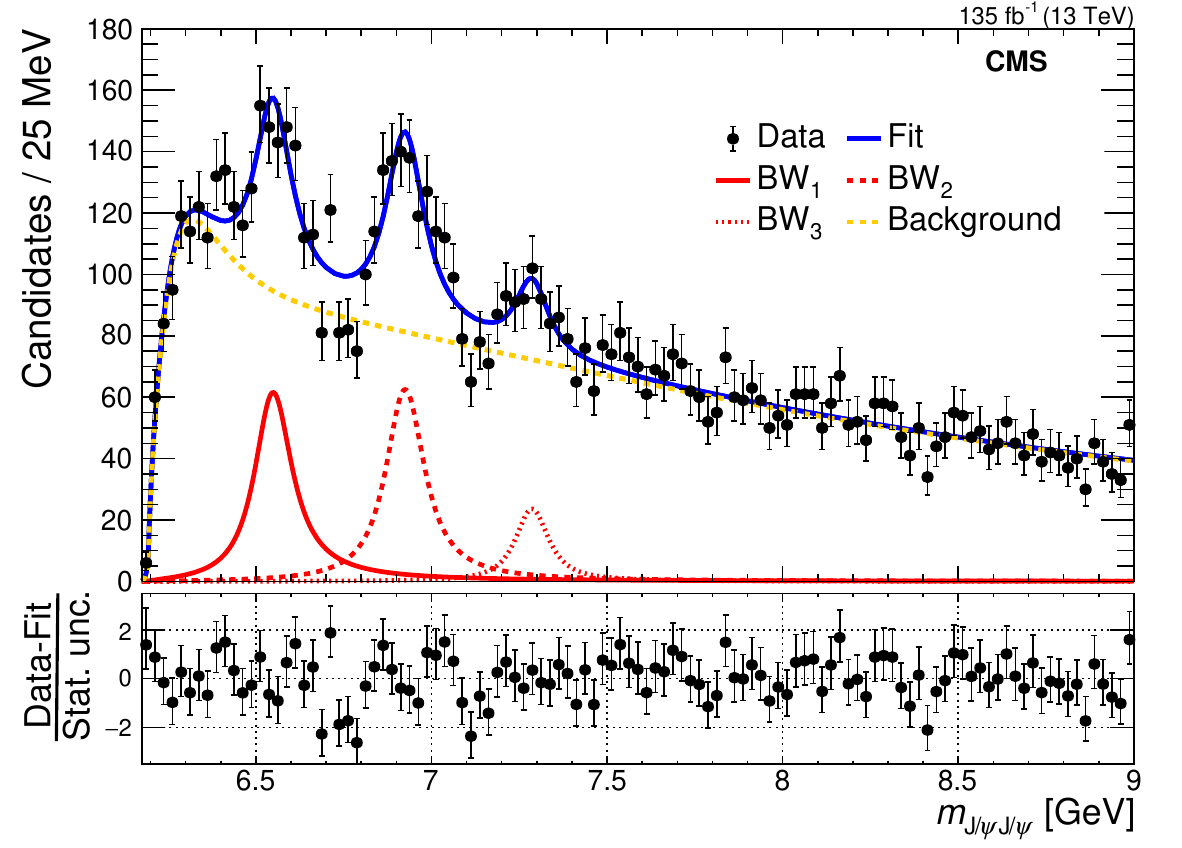}
\caption{Without interference}
\label{fig:cms-prl-nointerf}
\end{subfigure}\hfill
\begin{subfigure}[t]{0.48\textwidth}
\centering
\includegraphics[width=\linewidth]{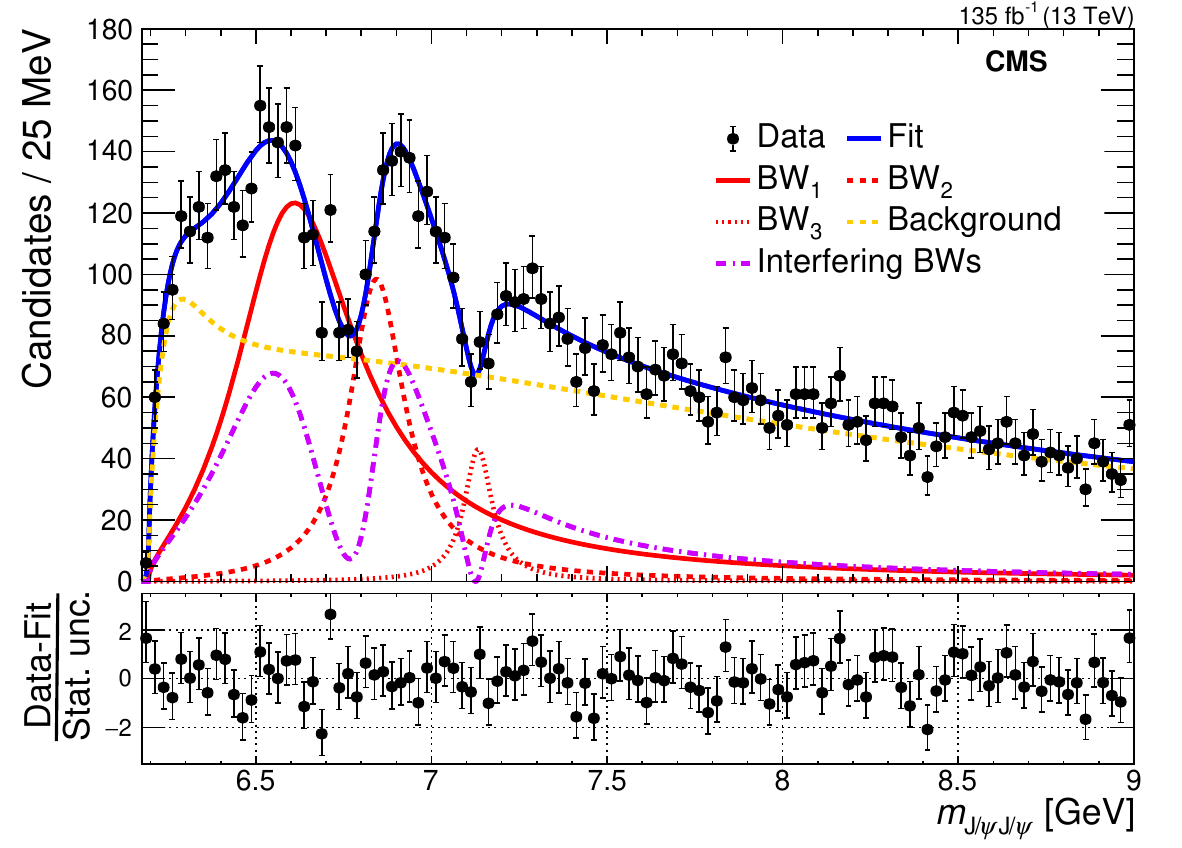}
\caption{With interference}
\label{fig:cms-prl-interf}
\end{subfigure}
\caption{$J/\psi\,J/\psi$ invariant-mass spectrum measured by CMS with the Run~2 dataset. Fits are shown for models (a)~without and (b)~with interference among three Breit--Wigner amplitudes~\cite{CMS:2023owd}.}
\label{fig:cms-prl}
\end{figure}

\begin{figure}[htbp]
\centering
\begin{subfigure}[t]{0.48\textwidth}
\centering
\includegraphics[width=0.8\linewidth]{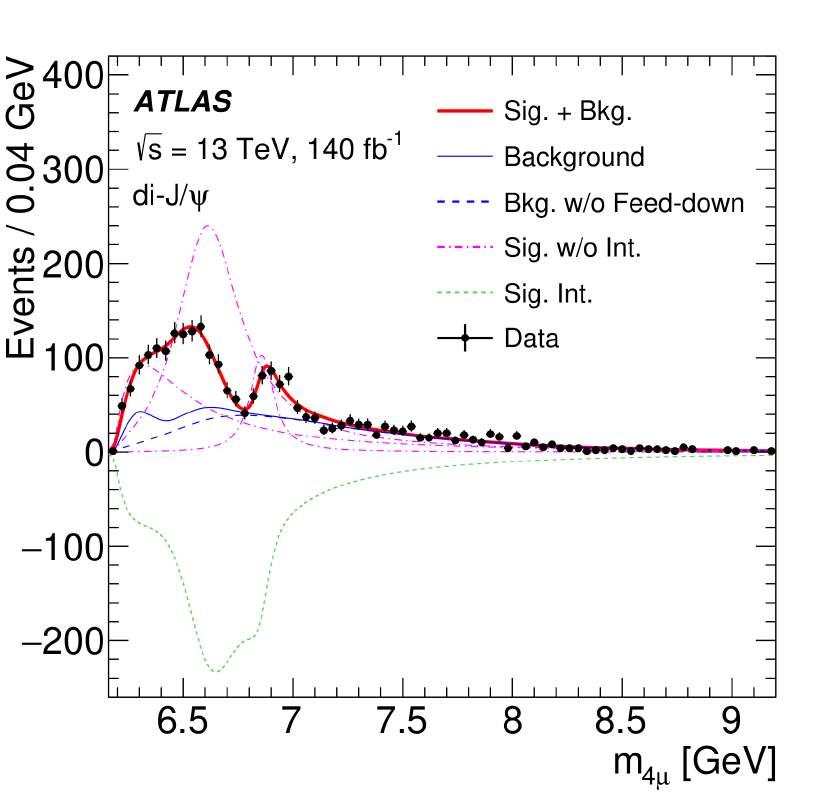}
\caption{Model A}
\label{fig:atlas-prl-A}
\end{subfigure}\hfill
\begin{subfigure}[t]{0.48\textwidth}
\centering
\includegraphics[width=0.8\linewidth]{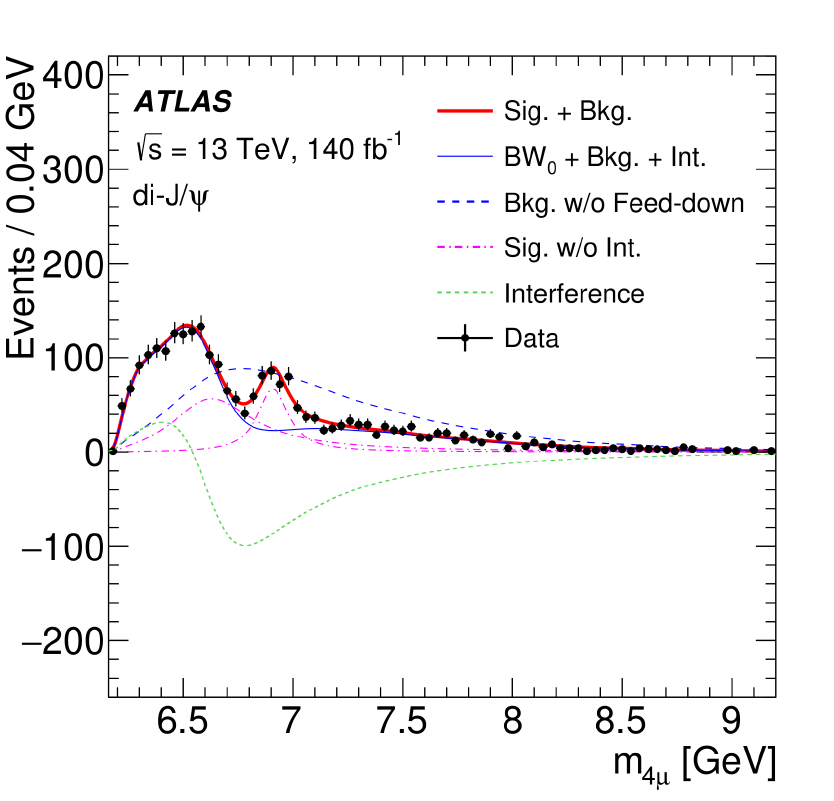}
\caption{Model B}
\label{fig:atlas-prl-B}
\end{subfigure}
\caption{$J/\psi\,J/\psi$ invariant-mass spectrum measured by ATLAS. Fits are shown for (a)~Model~A and (b)~Model~B, which incorporate different resonance and interference assumptions~\cite{ATLAS:2023bft}.}
\label{fig:atlas-prl}
\end{figure}

Applying this strategy to the CMS Run~2 $J/\psi\,J/\psi$ dataset, the Collaboration first reported three structures in a 2022 Physics Analysis Summary, based on a fit without interference~\cite{CMS:PAS-BPH-21-003}. The subsequent publication in 2024~\cite{CMS:2023owd} established $X(6600)$, $X(6900)$, and $X(7100)$ and showed that a global fit including interference among these states describes the $J/\psi\,J/\psi$ mass spectrum significantly better than a fit without interference (Fig.~\ref{fig:cms-prl}). The $X(6900)$ state, first reported by LHCb~(Fig.~\ref{fig:lhcb})~\cite{LHCb:2020bwg}, was confirmed independently by the ATLAS~(Fig.~\ref{fig:atlas-prl})~\cite{ATLAS:2023bft} and CMS~\cite{CMS:2023owd} Collaborations.

The observation of the $X(6600)$ state and evidence for the higher-mass $X(7100)$ suggested not an isolated resonance but a family of related states. LHCb had already shown that interference can be important for describing the $J/\psi\,J/\psi$ spectrum~\cite{LHCb:2020bwg}; the CMS fit included interference among the overlapping resonances themselves.

\begin{figure}[htbp]
\centering
\begin{subfigure}[t]{0.48\textwidth}
\centering
\includegraphics[width=\linewidth]{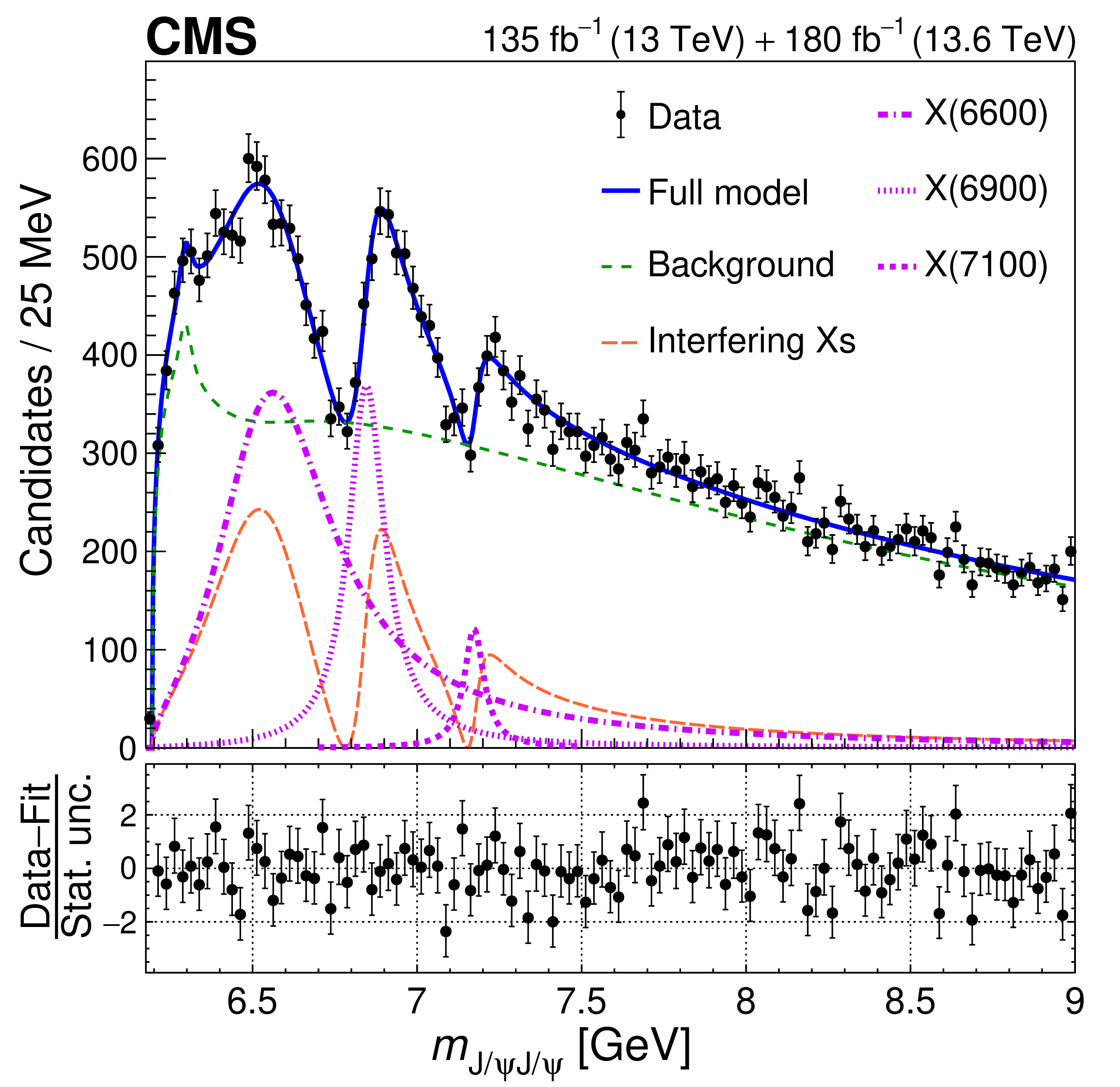}
\caption{$J/\psi\,J/\psi$}
\label{fig:cms-jj}
\end{subfigure}\hfill
\begin{subfigure}[t]{0.48\textwidth}
\centering
\includegraphics[width=\linewidth]{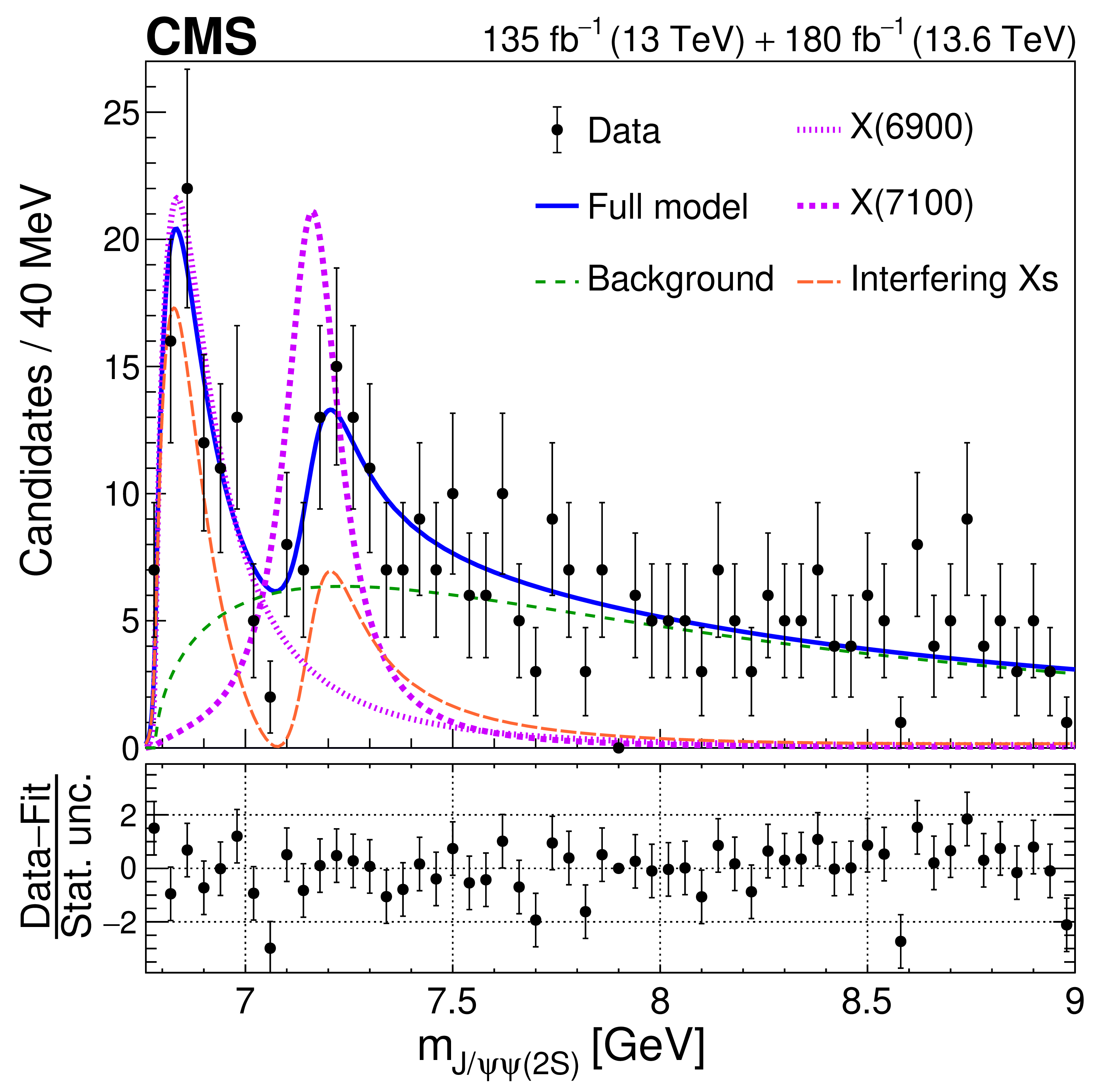}
\caption{$J/\psi\,\psi(2S)$}
\label{fig:cms-psi2s}
\end{subfigure}
\caption{Invariant-mass spectra from the combined Run~2 and Run~3 CMS dataset in the (a)~$J/\psi\,J/\psi$ and (b)~$J/\psi\,\psi(2S)$ channels. Both spectra are described by interference models~\cite{CMS:2025family}.}
\label{fig:cms-family}
\end{figure}

An analysis of the independent Run~3 $J/\psi\,J/\psi$ sample confirmed $X(6600)$, $X(6900)$, and $X(7100)$ and the interference among them with significances of well above $5\sigma$~\cite{CMS:2025family}. Combining the full Run~2 dataset with early Run~3 data, CMS accumulated approximately $315\,\mathrm{fb}^{-1}$, more than tripling the $J/\psi\,J/\psi$ candidates of the original analysis. The substantially increased statistics, together with improved modeling of the signal and background components, reduced the uncertainties on the resonance parameters by factors of two to three (Fig.~\ref{fig:cms-jj}). With the combined sample, interference among the resonances is preferred at more than $5\sigma$: an incoherent sum of independent Breit--Wigner peaks does not reproduce the dips between the structures.

\subsection{Independent Confirmation in the \texorpdfstring{$J/\psi\,\psi(2S)$}{J/psi psi(2S)} Channel}

With the combined Run~2 and Run~3 dataset, CMS observed the $X(6900)$ in the $J/\psi\,\psi(2S)$ final state with a significance of $8.1\sigma$ and found evidence for the $X(7100)$ at $4.3\sigma$ (Fig.~\ref{fig:cms-psi2s})~\cite{CMS:2025family}; the $X(6600)$ lies below the $J/\psi\,\psi(2S)$ threshold. Interference between $X(6900)$ and $X(7100)$ in this channel is corroborated at $2.5\sigma$. In the same channel, ATLAS had reported a $4.7\sigma$ excess near $6.9\,\mathrm{GeV}$ in the four-muon final state~\cite{ATLAS:2023bft}, and later, using both $\psi(2S)\to\mu^{+}\mu^{-}$ and $\psi(2S)\to J/\psi\pi^{+}\pi^{-}$, found an $8.9\sigma$ excess near $6.9\,\mathrm{GeV}$ and no significant signal near $7.2\,\mathrm{GeV}$~\cite{ATLAS:2025psi2s}.

\section{Determination of the Quantum Numbers}

If discovering a new resonance is the first step in hadron spectroscopy, determining its quantum numbers and measuring the cross section are the crucial next steps. Masses and widths alone cannot distinguish among competing interpretations such as compact tetraquarks, molecules, threshold effects, or kinematic enhancements. Angular distributions provide direct access to spin and parity, offering a powerful probe of the underlying structure.
Following the discovery of the fully charmed tetraquark family, CMS performed the first spin-parity analysis of these states using a multidimensional likelihood framework incorporating helicity amplitudes, detector effects, efficiency corrections, and resonance interference.

\begin{figure}[htbp]
\centering
\includegraphics[width=0.736\textwidth]{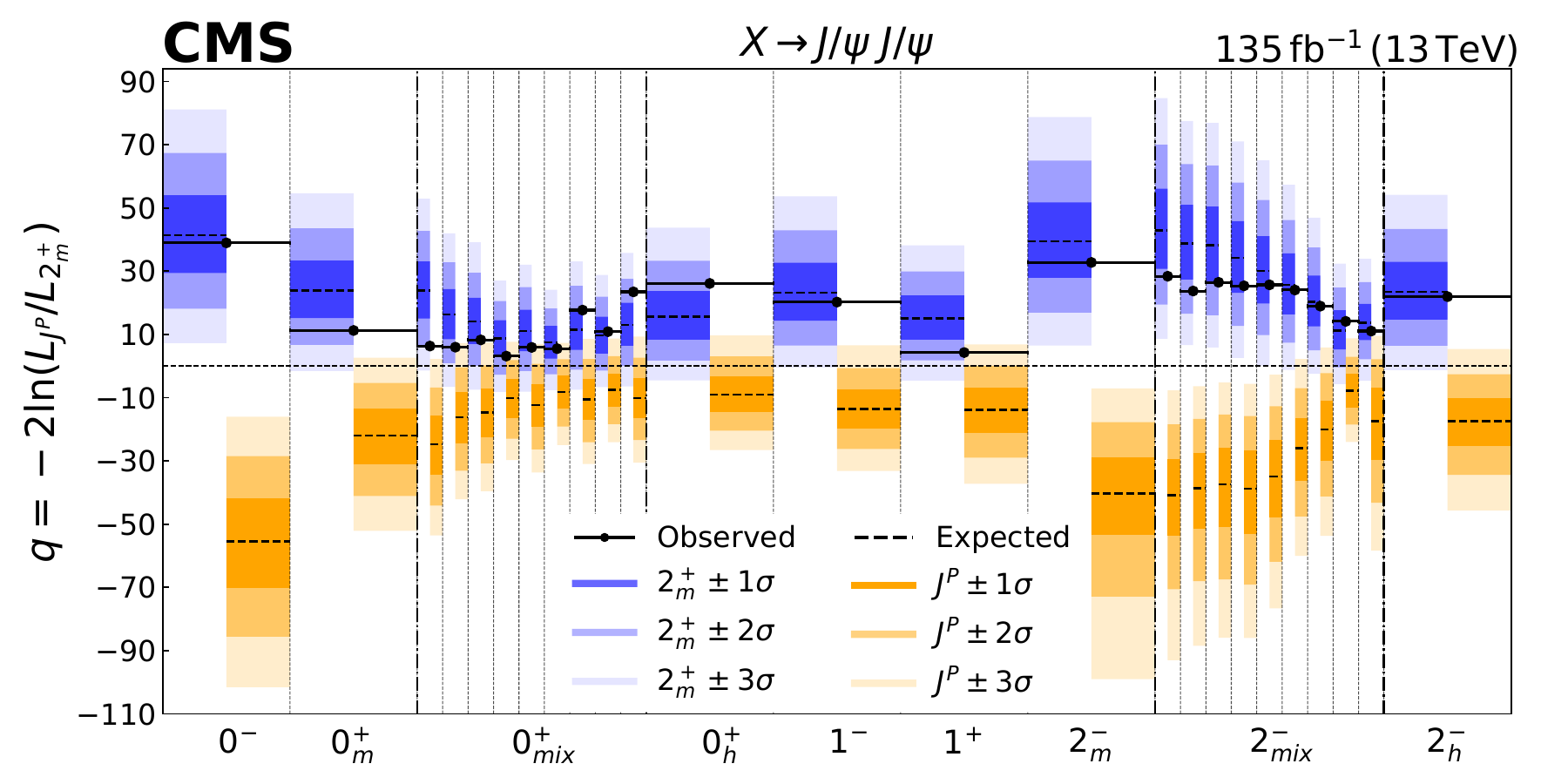}
\caption{Summary of spin-parity tests for the all-charm tetraquark family. Each alternative $J^{P}$ is compared with the favored $2^{+}_{m}$ assignment; colored bands mark the expected $1\sigma$, $2\sigma$, and $3\sigma$ ranges~\cite{CMS:2025fpt}.}
\label{fig:cms-jpc}
\end{figure}

Published in \textit{Nature} in 2025~\cite{CMS:2025fpt}, the analysis determined $J^{PC} = 2^{++}$ for the dominant state (Fig.~\ref{fig:cms-jpc}), marking the first experimental spin-parity measurement of a fully charmed tetraquark and the first for a promptly produced exotic hadron at the LHC.

\section{An Emerging Physical Picture}

Successive CMS measurements have transformed fully charmed tetraquark studies from isolated observations into an emerging spectroscopy~\cite{Zhu:2024swp,CMS:2025family,CMS:2025fpt,Wen:2026sspma}. The current picture reveals several structures between approximately $6.6$ and $7.1\,\mathrm{GeV}$. Whether they form a family of radial excitations, possibly of a tensor state, remains an open question.

\begin{figure}[htbp]
\centering
\includegraphics[width=0.50\textwidth]{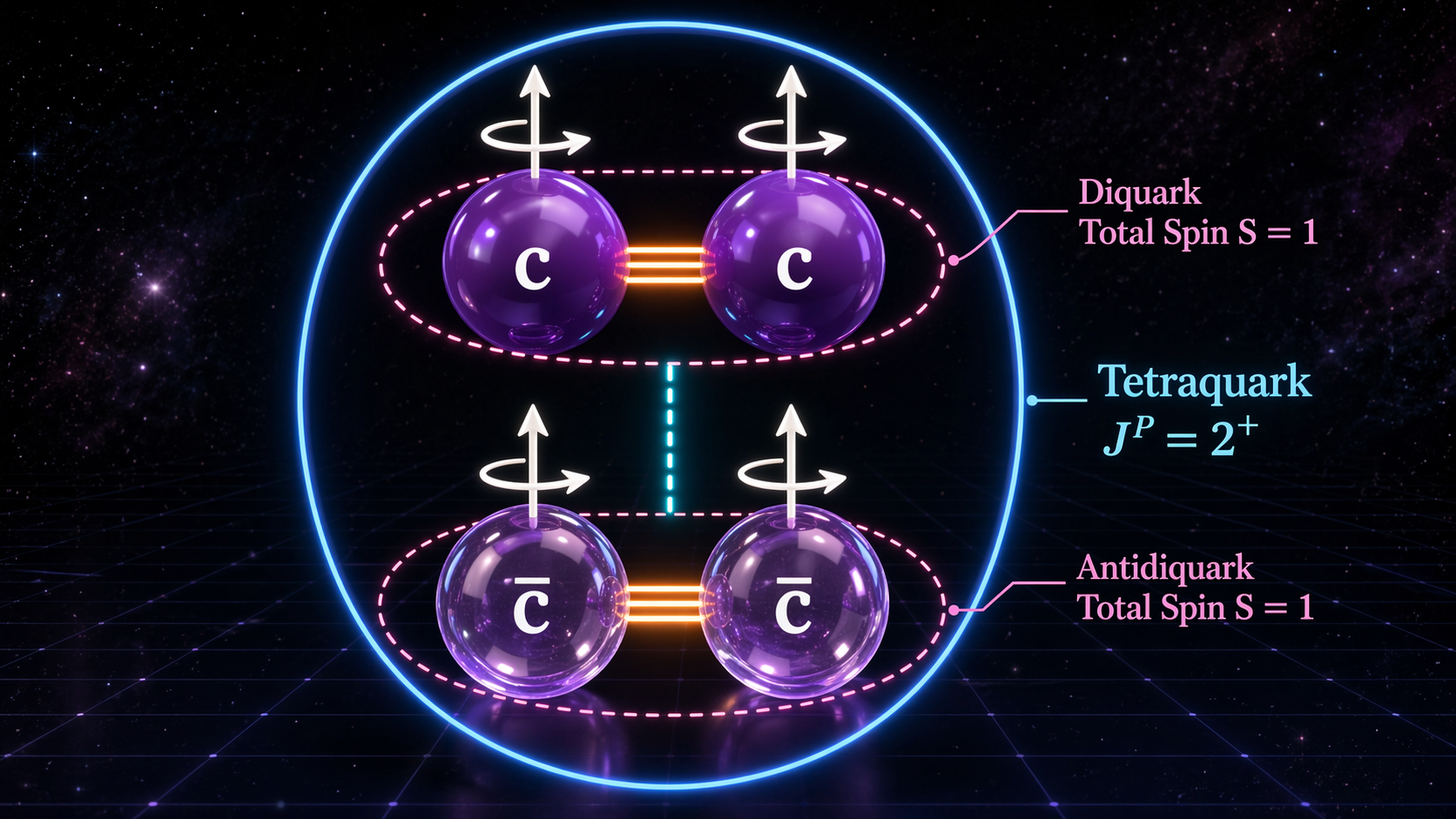}
\caption{Schematic of a diquark--antidiquark configuration with $J^{P}=2^{+}$: two charm quarks form a spin-1 diquark, two anticharm quarks form a spin-1 antidiquark, and the two couple to a tensor tetraquark~\cite{Wang:2026phd}.}
\label{fig:diquark}
\end{figure}

Although other models are not ruled out, including hadronic molecules~\cite{Lu:2023ccs}, hybrid (gluonic) configurations~\cite{Wan:2020fsk}, and coupled-channel dynamics near charmonium-pair thresholds~\cite{Zhou:2022xpd,Bai:2026atm}, the $J^{PC}=2^{++}$ assignment favors a diquark--antidiquark interpretation, in which a spin-1 $cc$ diquark and a spin-1 $\bar{c}\bar{c}$ antidiquark couple to a tensor state (Fig.~\ref{fig:diquark})~\cite{Debastiani:2017msn,Bedolla:2019uqp}. This marks a step forward in understanding the emerging picture, while a complete understanding still requires a global picture combining masses, widths, production rates, quantum numbers, decay patterns, polarization, and interference effects, much as in the charmonium spectroscopy of the 1970s.
Similar patterns are also seen in other vector--vector systems: three spin-2 states in the $\phi\phi$ spectrum~\cite{BESIII:2016cpx}, and a triplet of spin-1 states in the mixed-flavor $J/\psi\phi$ system~\cite{Zhou:2025nyj}. Studying these systems together with the all-charm family may lead to a better understanding of tetraquark spectroscopy.

\section{Summary and Outlook}

The history of fully charmed tetraquarks illustrates how major discoveries in particle physics often require decades of sustained theoretical and experimental effort. Nearly forty years after the first theoretical predictions~\cite{Iwasaki:1975pv,Ader:1981eb}, these systems were finally established experimentally at the LHC~\cite{LHCb:2020bwg,CMS:2023owd,CMS:2025family,CMS:2025fpt}.

CMS has played a central role in this progress, evolving from exploratory Run~1 studies into a systematic spectroscopy program. Its contributions include the observation of $X(6600)$ and $X(7100)$, the establishment of interference among the three resonances, and the first determination of their quantum numbers.

Many fundamental questions remain, including the production mechanisms, internal structure, and excitation spectrum of these states. Future measurements at ATLAS, CMS, and LHCb will further constrain production rates, polarization, decay modes, and interference effects, while the High-Luminosity LHC and future colliders such as FCC will enable precision spectroscopy and searches for heavier multiquark systems. After a long journey from prediction to discovery, fully heavy tetraquark physics is entering a new era, with the potential to become as important for understanding QCD as charmonium and bottomonium.

This Perspective reviews the contributions of CMS to this program. The effort originated from 
exploratory studies initiated by Kai Yi at the University of 
Iowa in 2013. In 2019, he established the NNU CMS group and, 
together with Zhen Hu, subsequently formed the NNU and THU 
teams at CMS that led the fully charmed tetraquark analyses 
described above.
Those analyses included the 2024 Run~2 $J/\psi\,J/\psi$ measurement~\cite{CMS:2023owd}, the 2025 determination of the quantum numbers~\cite{CMS:2025fpt}, and the ongoing combined Run~2 and Run~3 study~\cite{CMS:PAS-BPH-24-003}. The 2024 Run~2 analysis~\cite{CMS:2023owd} involved the postdocs Jingqing Zhang and Muhammad Ahmad, the PhD students Hongwei Wen and Xining Wang, and the undergraduates Jingjing Gu and Jiahua Chen, together with more than ten additional graduate and undergraduate students from both teams. The 2025 spin-parity analysis~\cite{CMS:2025fpt} involved Jingqing Zhang and Xining Wang.
As Tsinghua University marks a century of physics, the THU CMS team's role in this spectroscopy, together with NNU, illustrates how a decades-old theoretical idea became a new experimental frontier of QCD.

\section{Acknowledgments}

This work is partially supported by the Natural Science Foundation of China under Grants No.~12535004 and No.~W2443001, as well as the Ministry of Science and Technology of China under Grants No.~2023YFA1605804.

\bibliographystyle{unsrtnat}
\bibliography{refs}

\end{document}